\documentclass[11pt]{article}

\usepackage{palatino}
\usepackage{mathpazo}

\usepackage{braket}
\usepackage{amsfonts}
\usepackage{amssymb}
\usepackage{amsmath}
\usepackage{latexsym}
\usepackage{amsthm}
\usepackage{eepic}
\usepackage{sectsty}
\usepackage{hyperref}
\usepackage{gastex}

\hypersetup{pdfpagemode=UseNone}

\theoremstyle{definition}

\newcommand{\tinyspace}{\mspace{1mu}}

\newcommand{\norm}[1]{\left\lVert\tinyspace#1\tinyspace\right\rVert}
\newcommand{\abs}[1]{\left\lvert\tinyspace #1 \tinyspace\right\rvert}
\newcommand{\tr}{\operatorname{Tr}}

\newcommand{\fid}{\operatorname{F}}
\newcommand{\setft}[1]{\mathrm{#1}}
\newcommand{\lin}[1]{\setft{L}\left(#1\right)}
\newcommand{\density}[1]{\setft{D}\left(#1\right)}

\def\real{\mathbb{R}}

\newcommand{\reg}[1]{\mathsf{#1}}
\newcommand{\class}[1]{\textup{#1}}

\def\X{\mathcal{X}}
\def\Y{\mathcal{Y}}
\def\Z{\mathcal{Z}}
\def\A{\mathcal{A}}
\def\B{\mathcal{B}}

\def\S{\mathcal{S}}
\def\T{\mathcal{T}}

\begin{document}

\title{\bf\LARGE Coherent state exchange in multi-prover\\
  quantum interactive proof systems}

\author{
  Debbie Leung\footnote{%
    Institute for Quantum Computing and
    Department of Combinatorics and Optimization,
    University of Waterloo,
    Waterloo, Ontario, Canada.}
  \and
  Ben Toner\footnote{%
    School of Physics,
    University of Melbourne,
    Melbourne, Victoria, Australia.
    This work done while at Centrum Wiskunde \& Informatica, 
    Amsterdam, The Netherlands.}
  \and
  John Watrous\footnote{%
    Institute for Quantum Computing and
    School of Computer Science,
    University of Waterloo,
    Waterloo, Ontario, Canada.}
}

\date{February 28, 2011}

\maketitle

\begin{abstract}
  We show that any number of parties can coherently exchange any one
  pure quantum state for another, without communication, given prior
  shared entanglement.
  Two applications of this fact to the study of multi-prover quantum
  interactive proof systems are given.
  First, we prove that there exists a one-round two-prover quantum
  interactive proof system for which no finite amount of shared
  entanglement allows the provers to implement an optimal strategy.
  More specifically, for every fixed input string, there exists a
  sequence of strategies for the provers, with each strategy requiring
  more entanglement than the last, for which the probability for the
  provers to convince the verifier to accept approaches~1.
  It is not possible, however, for the provers to convince the
  verifier to accept with certainty with a finite amount of shared
  entanglement.
  The second application is a simple proof that multi-prover quantum
  interactive proofs can be transformed to have near-perfect
  completeness by the addition of one round of communication.
  
\end{abstract}

\section{Introduction}

The idea that entanglement may be used as a resource is central to the
theory of quantum communication and cryptography.
Well-known examples include teleportation \cite{BennettBCJPW93} and the
super-dense coding of both classical and quantum data
\cite{BennettW92,BennettSST02,HarrowHL04}.
In cryptography, entanglement is used not only in some implementations
of quantum key-distribution \cite{Ekert91}, but also as a mathematical
tool in security proofs \cite{LoC99,ShorP00} of quantum
key-distribution protocols not based on entanglement (such as
\cite{BennettB84}).
In these settings it may be said that the relationship between
entanglement and other resources (in particular quantum communication,
classical communication, and private shared randomness) is reasonably
well-understood \cite{BennettDSW96,DevetakHW03}.

There are, on the other hand, settings of interest where the
properties of entanglement as a resource are very poorly understood.
One example can be found in quantum communication complexity, wherein
it is not known if prior shared entanglement ever gives an asymptotic
reduction in the number of qubits of communication required to solve
general communication problems \cite{Wolf02,Brassard03}.
A second example, which is the main focus of this paper, concerns the
power of entanglement in the multi-prover interactive proof system
model, which has been studied in several recent papers 
\cite{
  KobayashiM03,
  CleveHTW04,
  Wehner06,
  CleveSUU08,
  CleveGJ09,
  KempeKMTV08,
  KempeRT08}.
Bell inequalities \cite{Bell64,ClauserH+69}, and many of the
open problems concerning them \cite{Gisin09}, have a fundamental
connection to this model (although not necessarily to our main
results).

Multi-prover interactive proof systems, which were first defined by
Ben-Or, Goldwasser, Kilian, and Wigderson \cite{Ben-OrGKW88}, involve
interactions among a {\it verifier} and two or more {\it provers}.
The verifier is always assumed to be efficiently implementable, while
the provers are typically permitted to have arbitrary complexity.
The verifier and provers each receive a copy of some input string $x$,
and then engage in an interaction based on this string.
During this interaction the verifier communicates privately with
each of the provers, possibly over the course of many rounds of
communication, but the provers are forbidden from communicating
directly with one another.
The provers may, however, agree on a joint strategy before the
interaction begins.

The provers act in collaboration to convince the verifier that the
input string $x$ is a {\it yes-input} to some fixed problem, and
therefore should be {\it accepted}.
The provers are not, however, considered to be trustworthy, and
so the verifier must be defined in such a way that it {\it rejects}
strings that are {\it no-inputs} to the problem being considered.
These two conditions---that the provers can convince the verifier to
accept yes-inputs but cannot convince the verifier to accept
no-inputs---are called the {\it completeness} and {\it soundness}
conditions, respectively, and are analogous to the notions in
mathematical logic that share these names.
In contrast to the notion of a mathematical proof, however, one
typically requires only that the completeness and soundness conditions
for interactive proof systems hold with high probability (for every
fixed yes or no input string).

The study of interactive proof systems, including single-prover and
multi-prover models, has had an enormous impact on the fields of
computational complexity and theoretical cryptography
\cite{AroraB09,Goldreich01}.
In particular, multi-prover interactive proof systems, and the
characterization of their expressive power \cite{BabaiFL91}, led
to the discovery of the 
{\it PCP} (or {\it Probabilistically Checkable Proof}) {\it Theorem}
\cite{AroraLMSS98,AroraS98,Dinur07}, which was a critical breakthrough
in understanding the hardness of approximation problems.

In the quantum setting, one must consider the possibility that the
provers initially share entanglement, which they might use as part of
their strategy during the interaction.
With this seemingly small change, nearly everything we know about
classical multi-prover interactive proof systems becomes invalid
within the quantum model.
The following points illustrate the effect of this change on our
current state of knowledge.

\begin{itemize}
\item
When the provers are not allowed to share prior entanglement, it is
known that the class of promise problems that have multi-prover
interactive proof systems is precisely \class{NEXP}, the class of
problems that can be solved nondeterministically in exponential time.
This holds for both classical \cite{BabaiFL91} and quantum
\cite{KobayashiM03} multi-prover interactive proofs.

\item
There are no nontrivial bounds known for the class of promise problems
having multi-prover interactive proofs when the provers initially
share entanglement, including the cases of both a classical and
quantum verifier.
At one extreme, such proof systems could have the same expressive
power as single-prover interactive proof systems, and at the other
extreme it is possible that uncomputable problems have such proof
systems.
\end{itemize}

One very basic question about multi-prover interactive proofs with
entangled provers has remained unanswered, and is closely related to
the lack of good upper bounds on their power:
{\it For a given verifier and input string, how much entanglement is
  needed for the provers to play optimally?}
To obtain an upper bound on the expressive power of multi-prover
interactive proofs with entangled provers, one seeks a general bound:
a limit on the amount of entanglement, as a function of the verifier's
description and the given input, needed for the provers to play
optimally.

Our first main result explains the difficulty in answering this
question: we prove that there exist two-prover quantum interactive
proof systems for which no finite amount of entanglement allows for an
optimal strategy on any fixed input.
In other words, there are interactive proofs such that, no matter what
entangled state the provers choose on a given input, it would always be
possible for them to do strictly better with more entanglement.
There is, therefore, no strict upper bound of the form discussed
above.
This fact has an obvious but important implication: one must consider
upper bounds on entanglement for {\it close-to-optimal} strategies if
this approach is to yield upper bounds on the power of multi-prover
quantum interactive proofs.

Our second main result concerns methods to achieve 
{\it perfect completeness} of quantum interactive proof systems while
retaining small soundness error.
In the single-prover case, an efficient transformation for doing this
exists that is both simple and easy to analyze \cite{KitaevW00}.
In the multi-prover setting, an analogous result was recently obtained
by Kempe, Kobayashi, Matsumoto, and Vidick \cite{KempeKMV09} based on
a much more complicated transformation and analysis.
This more complicated transformation was designed to handle the
locality constraints imposed on multiple provers.
It turns out, however, that this complicated procedure is not needed
after all, provided one is willing to make a small sacrifice.
We prove that the simple single-prover technique can be applied in the
multi-prover case to yield a proof system with {\it near-perfect}
completeness: honest provers are able to convince the verifier to
accept yes-inputs with any probability smaller than~1 that they
desire---but they might never reach probability 1 using finite
resources.
(For instance, the provers may choose to cause the verifier to accept
yes-inputs with probability $1 - \varepsilon(n)$ where
$\varepsilon(n)$ is the reciprocal of the busy beaver function.
The implementation of such a strategy, however, could require an
enormous amount of prior shared entanglement.)

The two main results just discussed are connected by the notion of
{\it coherent state exchange}, which is discussed in the next
section.
The first main result is then proved in Section~\ref{sec:game}, while
the second is proved in Section~\ref{sec:perfect-completeness}.
The paper concludes with Section~\ref{sec:conclusion}.
Throughout the paper we assume the reader is familiar with quantum
computing and with basic aspects of classical and quantum interactive
proof systems.
Our notation and terminology are consistent with other papers on these
topics.

\section{Coherent state exchange} \label{sec:exchange}

We begin by defining {\it coherent state exchange} as follows.  
Consider $m$ players $P_1,\ldots,P_m$; and suppose, for each
$i\in\{1,\ldots,m\}$, that player $P_i$ holds a quantum system whose
associated Hilbert space is denoted $\X^i$.  
The spaces $\X^i$ and $\X^j$ need not have equal dimension for
$i\not=j$.  
For two chosen pure states
$\ket{\phi},\ket{\psi}\in\X^1\otimes\cdots\otimes\X^m$,
we consider the situation in which the players wish to transform a
shared copy of $\ket{\phi}$ into $\ket{\psi}$, or vice versa.
We require that this task is completed 
{\it (1)~without communication} and 
{\it (2)~by a coherent process}.

To say that a process that performs state exchange is {\it coherent}
means that it can be applied in a way that preserves superpositions.
In particular, this means that it is possible for the players to
implement a transformation of the form
\[
\alpha\ket{0^m}\ket{\gamma}
+\beta\ket{1^m}\ket{\phi}
\mapsto
\alpha\ket{0^m}\ket{\gamma}
+\beta\ket{1^m}\ket{\psi},
\]
where the first $m$ qubits represent control qubits, with one held by
each player, and where the state
$\ket{\gamma}\in\X^1\otimes\cdots\otimes\X^m$ represents an arbitrary
state shared by the players.

In the absence of additional resources, it is not possible in general
to perform this task when $m \geq 2$.
In particular, given that the players cannot create entanglement out
of thin air, the task is easily seen to be impossible when the target
state $\ket{\psi}$ has more entanglement than the initial
state~$\ket{\phi}$.
However, if we consider the situation in which the players initially
share an auxiliary quantum state, and we allow this state to be
perturbed slightly by the process, then the above impossibility
argument based on entanglement is no longer valid---and as we will show,
the task indeed becomes possible.
We note that the coherence condition requires that a process of this
sort to leave the auxiliary quantum state nearly unchanged, in essence
using it as a catalyst.
The players cannot, for instance, simply swap the input state
$\ket{\phi}$ with an initially shared copy of $\ket{\psi}$ without
losing coherence.

For the case of $m=2$, one solution to this problem can be obtained
through the use of van~Dam and Hayden's 
{\it quantum state embezzlement}~\cite{vanDamH03}.
In quantum state embezzlement, two parties (Alice and Bob) perform a
transformation of the form $\ket{E_N} \mapsto \ket{E_N'}\ket{\phi}$,
for some shared entangled state $\ket{\phi}$ of their choice, where
$\{\ket{E_N}\}$ is a special family of states defined so that it
is possible to perform such a transformation in which $\ket{E_N}
\approx \ket{E_N'}$ for large $N$.  
Thus, they ``embezzle'' $\ket{\phi}$ from $\ket{E_N}$, leaving little
trace of their crime.
The process for doing this described by van Dam and Hayden is coherent
and requires no communication, and can therefore be done twice (once
in reverse) to achieve coherent state exchange.
It relies, however, on a representation of two-party pure quantum
states that no longer exists for $m$-party states when $m\geq 3$.

Here, we show that coherent state exchange for any number of parties
is always possible, with near-perfect coherence.
To simplify the description of the procedure, we will assume that
$\ket{\phi}$ and $\ket{\psi}$ are orthogonal.
(The more general case where $\ket{\phi}$ and $\ket{\psi}$ are not
necessarily orthogonal is easily handled, and is discussed in
Section~\ref{sec:non-orthogonal}.)

Let $N$ be a positive integer, which will determine the accuracy of
the procedure.
We assume that each player $P_i$ holds $N+2$ identical registers
labelled $\reg{X}_0^{i},\ldots,\reg{X}_{N+1}^{i}$, where each register
has an associated Hilbert space that is isomorphic to $\X^i$.
We take the initial state of the registers
$(\reg{X}^1_1,\ldots,\reg{X}^m_1),\ldots,
(\reg{X}^1_{N+1},\ldots,\reg{X}^m_{N+1})$
to be
\begin{equation} \label{eq:shared}
\ket{E_N} = \frac{1}{\sqrt{N}}
\sum_{k = 1}^N \ket{\phi}^{\otimes k} \ket{\psi}^{\otimes (N-k+1)},
\end{equation}
and we consider the case where the initial state of the registers
$(\reg{X}^1_0,\ldots,\reg{X}^m_0)$ is the input state~$\ket{\phi}$.
Thus, the state \eqref{eq:shared} represents the entanglement
initially shared by $P_1,\ldots,P_m$.
The procedure that transforms $\ket{\phi}$ into $\ket{\psi}$ is
simple: each player $P_i$ cyclically shifts the contents of the registers
$\reg{X}^i_0,\ldots,\reg{X}^i_{N+1}$ by applying a unitary operation
defined by the action
\[
\ket{x_0}\ket{x_1}\cdots\ket{x_{N+1}} \mapsto
\ket{x_{N+1}}\ket{x_0}\cdots\ket{x_N}
\]
on standard basis states.

Let us now consider the properties of the above procedure. 
It is clear that after the cyclic shift, the registers
$(\reg{X}^1_0,\ldots,\reg{X}^m_0)$ will contain a perfect copy of
$\ket{\psi}$, and the remaining registers will contain the state
\begin{equation} \label{eq:sharedafter}
\ket{E'_N} = \frac{1}{\sqrt{N}}
\sum_{k = 1}^N \ket{\phi}^{\otimes (k+1)} \ket{\psi}^{\otimes (N-k)}.
\end{equation}
Thus, the procedure transforms $\ket{\phi} \ket{E_N}$ into
$\ket{\psi} \ket{E_N'}$.
It is easily checked that $\braket{E'_N | E_N} = 1 - 1/N$, so for
large $N$ there is only a small disturbance to the shared entangled state
$\ket{E_N}$.
The procedure satisfies the coherence requirement for this reason.

The fidelity between $\ket{E_N}$ and $\ket{E_N'}$ can be
improved\footnote{This improvement was communicated to us by Aram
  Harrow, who attributes the idea to Peter Shor.}
to $1-O(1/N^2)$ if an alternative choice of the state $\ket{E_N}$ is made
in \eqref{eq:shared}.  
Here, the amplitude of the $k$-th term 
$\ket{\phi}^{\otimes k} \ket{\psi}^{\otimes (N-k+1)}$ is
changed from $\frac{1}{\sqrt{N}}$ to 
$a_k = \frac{\sqrt{2}}{\sqrt{N+1}} \sin(\frac{\pi k}{N{+}1})$.  
When $\ket{\phi}$ and $\ket{\psi}$ are orthogonal, the fidelity
$\braket{E'_N|E_N}$ is equal to 
$\sum_{k=2}^{N} a_k a_{k-1} = 1-O(1/N^2)$.  
(For arbitrary $\ket{\phi}$ and $\ket{\psi}$, a similar
result holds for the first coherent exchange protocol given later in
Section~\ref{sec:non-orthogonal} if the corresponding alternative
choices for $\ket{E_N}$ and $\ket{F_N}$ are used instead.)

In the case that the players wished instead to transform $\ket{\psi}$
to $\ket{\phi}$, so that the initial state of the registers
$(\reg{X}^1_0,\ldots,\reg{X}^m_0)$ is $\ket{\psi}$, the same state
\eqref{eq:shared} may be used, but the registers are shifted in the
opposite direction.

\subsection*{Further connections to embezzlement and other work}

A notion related to coherent state exchange, known as
{\it catalytic transformation} of pure states, was considered
by Jonathan and Plenio \cite{JonathanP99}.
In particular, they considered the situation in which two parties
transform one pure state to another (by local operations and classical
communication) using a catalyst---or a state that assists but is
left unchanged by the process.
Coherent state exchange (and embezzlement) do almost exactly this, and
without the exchange of classical information, but in the approximate
sense that the catalyst is permitted to change slightly.

As mentioned above, in the case $m=2$ one may use quantum state
embezzlement twice to implement coherent state exchange.
The family of states $\{\ket{E_N}\}$ defined in \cite{vanDamH03}
also has the added property of being {\it universal}, or independent
of the state $\ket{\phi}$ to be embezzled.
We note that it is possible  to use our method to give universal
embezzling families for all~$m$.
To define a universal embezzling family for any fixed $m$, we may
consider an $\epsilon$-net of states $\{\ket{\psi}\}$ in $N^m$
dimensions (for $\varepsilon = 1/N$, say), take
$\ket{\phi}=\ket{0^m}$, and define the embezzling state for each
$N$ to be the tensor product of all the states $\ket{E_N}$ ranging
over the $\varepsilon$-net.  
The embezzlement of a particular state is then performed in the
most straightforward way.
Unlike the families of van Dam and Hayden for the case $m=2$, our
method is highly inefficient, but nevertheless establishes that
universal embezzling families exist for all~$m$.

\section{Finite entanglement is suboptimal} \label{sec:game}

The purpose of this section is to prove the first main result of the
paper, which is that there exist two-prover quantum interactive proof
systems for which no finite amount of entanglement allows for an
optimal strategy on any fixed input.
It suffices to define a two-prover quantum interactive proof system
having no dependence on the input $x$; or, in simpler terms, to
consider a cooperative game played by two players and moderated by a
referee.
This type of {\it cooperative quantum game} represents a
generalization of the {\it non-local games} model of \cite{CleveHTW04},
where now the referee can send, receive, and process quantum
information.

To be more precise, and to aid in the exposition that follows, we
define two-player, one-round cooperative quantum games as follows:

\begin{enumerate}
\item
The referee prepares three quantum registers
$(\reg{R},\reg{S},\reg{T})$ in some chosen state, and then sends
$\reg{S}$ to Alice and $\reg{T}$ to Bob.

\item
Alice and Bob transform the registers $\reg{S}$ and $\reg{T}$ sent to
them however they choose, resulting in registers $\reg{A}$ and
$\reg{B}$ that are sent back to the referee.

\item
The referee performs a binary-valued measurement on the
registers $(\reg{R},\reg{A},\reg{B})$.
The outcome~1 means that Alice and Bob win, while the outcome 0 means
that they lose.
\end{enumerate}

\noindent
The restrictions on Alice and Bob are the same as for provers in a
quantum interactive proof system: they are not permitted to
communicate once the game begins, but may agree on a strategy
beforehand.
Such a strategy may include the sharing of an entangled state of their
own choosing, which they may use when transforming the registers sent
to them.
The complexity of the referee, which corresponds to the verifier in an
interactive proof system, is ignored given that we no longer consider
an input string.

\subsection*{Description of the game} \label{sec:game-description}

Consider the two-player cooperative quantum game that is determined by
the following specification of the referee:
\begin{enumerate}
\item
Let $\reg{R}$ be a qubit register and let $\reg{S}$ and
$\reg{T}$ be qutrit registers.
The referee initializes the registers $(\reg{R},\reg{S},\reg{T})$ to
the state 
\[
\frac{1}{\sqrt{2}}\ket{0}\ket{00} + \frac{1}{\sqrt{2}}\ket{1}\ket{\phi}
\]
where
\[
\ket{\phi} = \frac{1}{\sqrt{2}} \ket{11} + \frac{1}{\sqrt{2}}
\ket{22}.
\]
The registers $\reg{S}$ and $\reg{T}$ are sent to Alice and Bob,
respectively.

\item
The referee receives $\reg{A}$ from Alice and $\reg{B}$ from
Bob, where $\reg{A}$ and $\reg{B}$ are both single-qubit registers.
The triple $(\reg{R},\reg{A},\reg{B})$ is measured with respect to the
projective measurement $\{\Pi_0,\Pi_1\}$, where 
$\Pi_0 = I - \ket{\gamma}\bra{\gamma}$ and
$\Pi_1 = \ket{\gamma}\bra{\gamma}$,
for $\ket{\gamma} = (\ket{000} + \ket{111})\sqrt{2}$.
In accordance with the conventions stated above, the outcome 1 means
that Alice and Bob win while 0 means that they lose.
\end{enumerate}

The intuition behind this game is as follows.
Alice and Bob are presented with two possibilities, in superposition:
they receive either the unentangled state $\ket{00}$ or the entangled
state $\ket{\phi}$. 
Their goal is essentially to do nothing to $\ket{00}$ and to convert
$\ket{\phi}$ to $\ket{11}$, for they want the referee to hold the
state $\ket{\gamma}$ when the final measurement is made.
These transformations must be done coherently, without measurements or
residual evidence of which of the two transformations
$\ket{00}\mapsto\ket{00}$ or $\ket{\phi} \mapsto \ket{11}$ was
performed, for otherwise the final state of the referee will not have
a large overlap with $\ket{\gamma}$.

The required transformation will be possible using coherent state
exchange, with a winning probability approaching 1. 
It will be shown, however, that it is never possible for Alice and Bob
to win with certainty, provided they initially share a finite
entangled state.

\subsection*{Strategies that win with probability approaching 1}

We now present a family of strategies for Alice and Bob that win with
probability approaching~1.
In the above game, Alice receives $\reg{S}$ from the referee
and returns $\reg{A}$; and likewise for Bob with registers
$\reg{T}$ and $\reg{B}$.
Alice will begin with the qubit $\reg{A}$ initialized to $\ket{0}$, and
Bob begins with $\reg{B}$ initialized to $\ket{0}$ as well.
Let $U$ be a unitary operation, acting on a pair consisting of a qutrit
and a qubit, with the following behavior:
\[
U: \ket{0}\ket{0} \mapsto \ket{0}\ket{0},\quad\quad
U: \ket{1}\ket{0} \mapsto \ket{1}\ket{1}, \quad\quad\text{and}\quad\quad
U: \ket{2}\ket{0} \mapsto \ket{2}\ket{1}.
\]
Upon receiving $\reg{S}$, Alice applies $U$ to $(\reg{S},\reg{A})$,
and Bob does likewise to $(\reg{T}, \reg{B})$ after receiving
$\reg{T}$.
This leaves the 5-tuple $(\reg{R},\reg{A},\reg{B},\reg{S},\reg{T})$
in the state
\[
\frac{1}{\sqrt{2}} \ket{000} \ket{00} +
\frac{1}{\sqrt{2}} \ket{111} \ket{\phi}.
\]

Alice and Bob have not yet sent $\reg{A}$ and $\reg{B}$ to the
referee.
Before sending these registers, they use them as control qubits to
transform registers $\reg{S}$ and $\reg{T}$ in superposition: if
$\reg{A}$ and $\reg{B}$ are set to $0$ then nothing happens, while if
$\reg{A}$ and $\reg{B}$ are 1 they perform coherent state
exchange, transforming $\ket{\phi}$ to $\ket{00}$.
Assuming that Alice and Bob initially share the entangled state
\[
\ket{E_N} = \frac{1}{\sqrt{N}}\sum_{k = 1}^N 
\ket{\phi}^{\otimes k}\ket{00}^{\otimes (N-k+1)},
\]
the resulting state is
\[
\frac{1}{\sqrt{2}} \ket{000} \ket{00}\ket{E_N} +
\frac{1}{\sqrt{2}} \ket{111} \ket{00}\ket{E_N'}
\]
for
\[
\ket{E'_N} = \frac{1}{\sqrt{N}}\sum_{k = 1}^N 
\ket{\phi}^{\otimes (k+1)}\ket{00}^{\otimes (N-k)}.
\]

The registers $\reg{A}$ and $\reg{B}$ are now sent to the referee,
whose measurement results in outcome 1 with probability
\[
\norm{
\frac{1}{2} \ket{E_N} + \frac{1}{2} \ket{E_N'}}^2 = 1 - \frac{1}{2N}.
\]

\subsection*{Impossibility to win with certainty}

Now we will prove that Alice and Bob cannot win with certainty
regardless of the strategy they employ.
Without loss of generality, it may be assumed that Alice and Bob
initially share a pure entangled state $\ket{\psi}\in\X_A\otimes\X_B$,
where the spaces $\X_A$ and $\X_B$ have the same dimension $d$.
When Alice and Bob receive $\reg{S}$ and $\reg{T}$ from the referee,
the state of the entire system is given by
\[
\frac{1}{\sqrt{2}}\ket{0}\ket{00}\ket{\psi} +
\frac{1}{\sqrt{2}}\ket{1}\ket{\phi}\ket{\psi}.
\]

General quantum operations performed by Alice and Bob can be
described by linear isometries: $A$ for Alice and $B$ for Bob.
These isometries take the form
$A: \S \otimes \X_A \rightarrow \A \otimes \Y_A$ and
$B: \T \otimes \X_B \rightarrow \B \otimes \Y_B$,
where $\S,\T,\A,\B$ are the spaces associated with the registers
$\reg{S}$, $\reg{T}$, $\reg{A}$, and $\reg{B}$, and the spaces $\Y_A$
and $\Y_B$ are arbitrary.
The state of the system immediately before the referee measures is
therefore
\[
\frac{1}{\sqrt{2}} \ket{0} (A\otimes B) \ket{00}\ket{\psi}
+\frac{1}{\sqrt{2}} \ket{1} (A\otimes B)\ket{\phi}\ket{\psi}.
\]
By defining operators $A_0,A_1\in\lin{\S\otimes\X_A,\Y_A}$ and
$B_0,B_1\in\lin{\T\otimes\X_B,\Y_B}$ as
\[
A_0 = \left(\bra{0} \otimes I\right) A, \quad
A_1 = \left(\bra{1} \otimes I\right) A, \quad
B_0 = \left(\bra{0} \otimes I\right) B,\quad
B_1 = \left(\bra{1} \otimes I\right) B,
\]
we may express the probability that Alice and Bob win as
\[
\frac{1}{4}\norm{
(A_0\otimes B_0) \ket{00}\ket{\psi}
+(A_1\otimes B_1) \ket{\phi}\ket{\psi}}^2
\leq
\frac{1}{2} + \frac{1}{2}
\abs{\bra{\phi} \bra{\psi} 
\left(A_1^{\ast}A_0\otimes B_1^{\ast}B_0\right)
\ket{00} \ket{\psi}}.
\]

We have that $\norm{A_1^{\ast}A_0} \leq 1$ and
$\norm{B_1^{\ast}B_0} \leq 1$, and therefore it is possible to express
both $A_1^{\ast}A_0$ and $B_1^{\ast}B_0$ as convex combinations of
unitary operators.
By convexity, the winning probability is therefore bounded above by
\[
\frac{1}{2} + \frac{1}{2}
\abs{\bra{\phi} \bra{\psi} (U_A \otimes U_B) \ket{00} \ket{\psi}}
\]
for some choice of unitary operators $U_A$ and $U_B$.
Notice that $(U_A\otimes U_B) \ket{00}\ket{\psi}$
must have 1 e-bit of entanglement less than $\ket{\phi}\ket{\psi}$,
and so the states cannot be equal---and therefore the success
probability cannot be 1.

A quantitative bound may be proved as follows.
Using one of the Fuchs--van de Graaf inequalities \cite{FuchsvdG99}
and the monotonicity of the fidelity under partial tracing, it holds
that
\[
\abs{\braket{\phi,\psi|U_A \otimes U_B|00,\psi}}
\leq \fid(\rho,\xi) \leq \sqrt{1 - \frac{1}{4}\norm{\rho - \xi}_1^2}
\]
for
\begin{align*}
\rho & = \tr_{\B\otimes\X_B} 
\left(\ket{\phi}\bra{\phi} \otimes \ket{\psi}\bra{\psi}\right),\\
\xi & = \tr_{\B\otimes\X_B}
U_A
\left(\ket{00}\bra{00} \otimes \ket{\psi}\bra{\psi}\right) U_A^{\ast}.
\end{align*}
At this point we may follow precisely the analysis of
van Dam and Hayden \cite{vanDamH03} for the near-optimality of their
universal embezzling families: we have that $S(\rho)-S(\xi)=1$,
from which it follows that
$\norm{\rho - \xi}_1 \geq 1/(2\log(3d))$
by Fannes' inequality \cite{Fannes73}.
Consequently, the winning probability is bounded above by
$1 - 1/(32 \log^2 (3d))$.
The error probability therefore decreases at most quadratically in the
number of qubits that Alice and Bob initially share.
%
%
Incidentally, this is saturated by the alternative choice of 
$\ket{E_N}$ described towards the end of Section \ref{sec:exchange}.  

\subsection*{Consequences for entanglement assisted local quantum
  operations}

\begin{figure}[t]
  \begin{center}
    \small
    \unitlength=0.65pt
    \begin{picture}(300, 200)(0,-20)
      \node[Nw=40,Nh=70,Nmr=3](A)(150,130){$\Psi_A$}
      \node[Nw=40,Nh=70,Nmr=3](B)(150,10){$\Psi_B$}
      \node[Nframe=n](rho)(100,70){$\rho$}
      \node[Nframe=n,Nw=40](xi)(-20,70){$\xi$}
      \node[Nframe=n,Nw=50](Out)(320,70){$\Phi(\xi)$}
      \drawbpedge[eyo=-10,ELpos=30](rho,90,40,A,180,40){$\Z_A$} 
      \drawbpedge[eyo=10,ELside=r,ELpos=30](rho,-90,40,B,180,40){$\Z_B$} 
      \drawbpedge[syo=5,eyo=10,ELpos=70](xi,0,150,A,180,150){$\X_A$} 
      \drawbpedge[syo=-5,eyo=-10,ELside=r,ELpos=70](xi,0,150,B,180,150){$\X_B$}
      \drawbpedge[eyo=5,ELpos=30](A,0,150,Out,180,150){$\Y_A$} 
      \drawbpedge[eyo=-5,ELside=r,ELpos=30](B,0,150,Out,180,150){$\Y_B$} 
    \end{picture}
  \end{center}
  \caption{An entanglement-assisted local quantum operation~$\Phi$.
    An input state $\xi \in \density{\X_A\otimes\X_B}$ is transformed into
    the output state $\Phi(\xi)$ by means of local quantum operations
    $\Psi_A$ and $\Psi_B$, along with a shared entangled state
    $\rho\in\density{\Z_A\otimes\Z_B}$.} 
  \label{fig:ealo}
\end{figure}
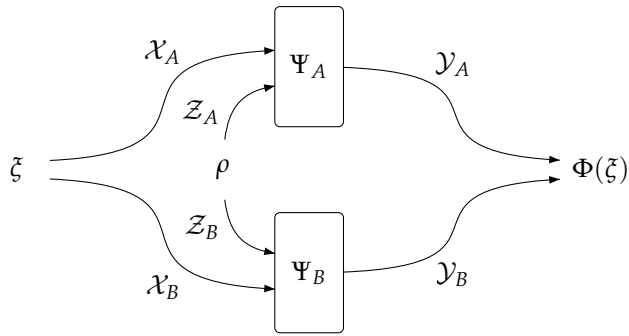

For fixed spaces $\X_A$, $\X_B$, $\Y_A$, and $\Y_B$, a quantum
operation
$\Phi: \lin{\X_A\otimes\X_B} \rightarrow \lin{\Y_A\otimes\Y_B}$
is an {\it entanglement assisted local quantum operation} if it can be
realized as illustrated in Figure~\ref{fig:ealo}; or more precisely,
if there exists some choice of spaces $\Z_A$ and $\Z_B$, a density
operator $\rho\in\density{\Z_A\otimes\Z_B}$, and admissible
super-operators $\Psi_A : \lin{\X_A\otimes\Z_A}\rightarrow\lin{\Y_A}$
and
$\Psi_B : \lin{\X_B\otimes\Z_B}\rightarrow\lin{\Y_B}$ 
such that
$\Phi(\xi) = (\Psi_A\otimes\Psi_B)(\rho\otimes\xi)$
for all $\xi\in\lin{\X_A\otimes\X_B}$.
Operations of this type are also known as {\it localizable}
operations \cite{BeckmanGNP01}.
In addition to having an obvious relevance to two-prover quantum
interactive proof systems, this is an interesting and fundamental class
of quantum operations in its own right.

An unfortunate fact that follows from the analysis of the game
presented above is the following.
When $\X_A$ and $\X_B$ have dimension at least 3 and $\Y_A$ and $\Y_B$
have dimension at least 2, the set of entanglement-assisted local
quantum operations 
$\Phi:\lin{\X_A\otimes\X_B} \rightarrow \lin{\Y_A\otimes\Y_B}$ is not
a closed set: the sequence of entanglement-assisted local quantum
operations induced by the strategies described above converges to a
valid quantum operation that is not an entanglement-assisted local quantum
operation.

\subsection*{Another connection with prior work}

We wish to point out one further connection between the above result and
some existing work.
In the exact catalytic transformation setting of Jonathan and Plenio
\cite{JonathanP99}, Daftuar and Klimesh \cite{DaftuarK01} proved the
following fact: the dimension of the catalyst required to transform
one state to another, when this is possible, cannot be bounded by any
function of the dimension of those states.
Although this fact does not have a direct implication to the
cooperative quantum games model, and is incomparable to our result as
far as we can see, there is a similarity in spirit between the results
that is worthy of note.

\section{Near-perfect completeness} \label{sec:perfect-completeness}

Kempe, Kobayashi, Matsumoto, and Vidick \cite{KempeKMV09} proved that
multi-prover quantum interactive proof systems can be efficiently
transformed to have {\it perfect completeness}, while retaining small
soundness error.
An analogous fact was previously shown to hold for single-prover
quantum interactive proof systems \cite{KitaevW00}, but the two proofs
are quite different.
The proof in \cite{KitaevW00} for the single-prover case is very
simple while the proof in \cite{KempeKMV09} for the multi-prover case
is rather complicated.
In this section we show that the use of coherent state exchange
allows the simple proof for the single-prover setting to be applied in
the multi-prover setting.

There is, however, one small caveat: whereas 
Kempe, Kobayashi, Matsumoto, and Vidick achieve truly perfect
completeness (in as far as quantum operations can ever be implemented
perfectly), we must settle for {\it near-perfect} completeness:
similar to the game from the previous section, honest provers will be
able to convince the verifier to accept yes-inputs with any
probability smaller than 1 that they desire, but the probability may
not in actuality be 1.
For most intents and purposes, though, we believe that this behavior
can reasonably be viewed as representing perfect completeness.

Suppose that a verifier $V$ interacts with $m$ provers
$P_1,\ldots,P_m$ for $r$ rounds, and suppose the completeness and
soundness probabilities for this verifier are given by $c$ and $s$,
respectively (which may be functions of the input length).
Specifically, for the promise problem 
$A = (A_{\text{yes}},A_{\text{no}})$ of interest, the following
conditions hold:

\begin{enumerate}
\item 
  Completeness.  The verifier is convinced to accept every
  yes-input $x\in A_{\text{yes}}$ with probability at least
  $c(\abs{x})$ by the provers' strategy.

\item
  Soundness. The verifier cannot be convinced to accept any no-input
  $x\in A_{\text{no}}$ with probability exceeding $s(\abs{x})$,
  regardless of the provers' strategy.
\end{enumerate}

\noindent
As usual and without loss of generality, we may ``purify'' a given
proof system so that the verifier $V$ and provers $P_1,\ldots,P_m$ are
described by unitary operations and the provers' initial shared
entanglement is pure.
We also make two simple assumptions on the proof system and the
completeness probability $c(\abs{x})$.
First, we assume that it is possible for the provers to convince the
verifier to accept every string $x\in A_{\text{yes}}$ with probability
exactly $c(\abs{x})$.
This can be achieved, for example, by appending an extra bit to 
the last message of the first prover and having the verifier reject
when this bit is 1.
Second, we assume that the value $c(\abs{x})$ is such that the
verifier can efficiently implement the rotation
\[
\ket{0} \mapsto \sqrt{1 - c(\abs{x})} \ket{0} -
\sqrt{c(\abs{x})} \ket{1},\quad\quad
\ket{1} \mapsto \sqrt{c(\abs{x})} \ket{0} +
\sqrt{1 - c(\abs{x})} \ket{1}
\]
without error.
(We also assume reversible computations incur no error.)

Now, assume that an input string 
$x\in A_{\text{yes}}\cup A_{\text{no}}$ has been fixed.
(As $x$ is now fixed, we will not explicitly refer to $x$ or $\abs{x}$
when discussing quantities depending on $x$.)
Let $p$ denote the probability that the verifier accepts.
Given the purity assumption of the proof system, this means that the
final state of the entire system at the end of the interaction may be
expressed as
\[
\sqrt{1-p} \ket{0}\ket{\phi_0} + \sqrt{p} \ket{1}\ket{\phi_1},
\]
where the first qubit in this expression represents the verifier's
output qubit.
The remaining part of the state, represented by $\ket{\phi_0}$ and
$\ket{\phi_1}$, corresponds to the state of every other register in
the proof system, shared in some arbitrary way among the verifier and
provers.
For simplicity we will assume that $\ket{\phi_0}$ and $\ket{\phi_1}$
are orthogonal, which at most requires that the verifier makes a
pseudo-copy of the output qubit as its last action.

To transform the proof system to one with near-perfect completeness,
one additional round of communication is added to the end of the
protocol.
To describe what happens in this additional round of communication,
let us write $\reg{A}$ to denote the verifier's output qubit,
$\reg{V}$ to denote the register comprising all of the verifier's
memory aside from the output qubit, and $\reg{P}_1,\ldots,\reg{P}_m$
to denote registers representing the provers' memories, all
corresponding to the final state of the original protocol.

To start the additional round of communication, the verifier prepares
$m$ additional single-qubit registers $\reg{A}_1,\ldots,\reg{A}_m$ as
pseudo-copies of $\reg{A}$, so that the state of the system becomes
\[
\sqrt{1-p} \ket{0}\ket{0^m}\ket{\phi_0} 
+ \sqrt{p} \ket{1}\ket{1^m}\ket{\phi_1}.
\]
The verifier then sends $\reg{V}$ to the first prover $P_1$ (which is
an arbitrary choice, but one that all provers are aware of), and sends
each register $\reg{A}_i$ to prover $P_i$.

Upon receiving these registers from the verifier, the provers perform
the following actions.
First, using the registers $(\reg{A}_1,\ldots,\reg{A}_m)$ as control
qubits, the provers perform coherent state exchange: when each
register $\reg{A}_i$ contains 0, nothing happens;
and when each register $\reg{A}_i$ contains~1, the state
$\ket{\phi_1}$ is exchanged for $\ket{\phi_0}$.
The resulting state of the entire system is
\[
\sqrt{1-p} \ket{0}\ket{0^m}\ket{\phi_0}\ket{E_N}
+ \sqrt{p} \ket{1}\ket{1^m}\ket{\phi_0}\ket{E_N'},
\]
where
\[
\ket{E_N} = \frac{1}{\sqrt{N}}
\sum_{k = 1}^N \ket{\phi_0}^{\otimes k} 
\ket{\phi_1}^{\otimes (N-k+1)}
\]
is an additional shared entangled state the provers use for this
purpose, and $\ket{E_N'}$ is defined in the same way as in
Section~\ref{sec:exchange}.
(This expression of $\ket{E_N}$ makes use of the assumption that
$\ket{\phi_0}$ and $\ket{\phi_1}$ are orthogonal.
One may instead consult the discussion in
Section~\ref{sec:non-orthogonal}, which does not require this
assumption.)
The number $N$ is the provers' choice for an accuracy parameter, which
we assume to be as large as they wish.
Once this is done, the provers return the registers
$\reg{A}_1,\ldots,\reg{A}_m$ to the verifier.

The final step is that the verifier measures the registers
$(\reg{A},\reg{A}_1,\ldots,\reg{A}_m)$ with respect to a basis
containing the state
$\sqrt{1 - c}\ket{0}\ket{0^m} + \sqrt{c}\ket{1}\ket{1^m}$,
accepting if the output matches this state.
(This is possible given our assumptions on $c$.)
In the case that $x\in A_{\text{yes}}$ the provers may take $p = c$,
and so the acceptance probability is
\[
\norm{(1 - c) \ket{E_N} + c \ket{E'_N}}^2 \geq 1 - \frac{1}{2N}.
\]
This is arbitrarily close to $1$, given that the provers may take any
value for $N$.
In the case that $x\in A_{\text{no}}$ we have $p\leq s$, from which it
is routine to show that the acceptance probability is at most
\[
\left(\sqrt{s}\sqrt{c} + \sqrt{1-s}\sqrt{1-c}\right)^2
\leq 1 - (c - s)^2.
\]

\section{Coherent exchange of non-orthogonal states}
\label{sec:non-orthogonal}

Here we briefly discuss coherent state exchange for non-orthogonal
states $\ket{\phi}$ and $\ket{\psi}$.
The simplest method that we have considered requires that
\[
\dim(\X^1 \otimes \cdots \otimes \X^m) \geq 3,
\]
which is immediate provided that $m\geq 2$ and that each $\X^i$ is
non-trivial.
One may then choose any state
$\ket{\eta}\in\X^1\otimes\cdots\otimes\X^m$ that is orthogonal to both
$\ket{\phi}$ and $\ket{\psi}$, and perform two state exchanges:
first from $\ket{\phi}$ and $\ket{\eta}$ and then from $\ket{\eta}$ to
$\ket{\psi}$.
The auxiliary state naturally takes the form $\ket{E_N}\ket{F_N}$,
where $\ket{E_N}$ is used to transform $\ket{\phi}$ to $\ket{\eta}$
and $\ket{F_N}$ is used to transform $\ket{\eta}$ to $\ket{\psi}$.
Aside from this change, no new analysis is required.

Another method is as follows.
Suppose 
$\braket{\phi|\psi} = a e^{i\theta}$ for $a>0$, and define
$\ket{\tilde{\psi}} = e^{-i\theta}\ket{\psi}$.
It is easy to coherently exchange $\ket{\tilde{\psi}}$ for
$\ket{\psi}$ by letting one player induce a global phase (which
translates into a phase shift on a control qubit if the process is
performed in superposition).
Thus, it remains to exchange $\ket{\phi}$ for $\ket{\tilde{\psi}}$.
If $a=1$ there is nothing to do, while if $a<1$ this may be done in a
similar way to the orthogonal case, through the use of the state
\[
\ket{E_N} = \frac{1}{\sqrt{N_1}}
\sum_{k = 1}^N \ket{\phi}^{\otimes k} \ket{\tilde{\psi}}^{\otimes (N-k+1)}.
\]
The only difference between this state and the one in
\eqref{eq:shared} is the normalization---it is obvious that $N\leq
N_1\leq N^2$, and more explicitly we have
\[
N_1 = \frac{1+a}{1-a} N - 2 a \frac{1-a^N}{(1-a)^2}.
\]
For 
\[
\ket{E'_N} = \frac{1}{\sqrt{N_1}}
\sum_{k = 1}^N \ket{\phi}^{\otimes k+1} \ket{\tilde{\psi}}^{\otimes (N-k)}
\]
we have
\[
\braket{E'_N | E_N} = 1 - \frac{1 - a^N}{N_1} \leq 1 - \frac{1}{N}.
\]
As is not surprising, the efficiency is no worse than in the
orthogonal case.

\section{Conclusion} \label{sec:conclusion}

We have discussed two applications of coherent state exchange to the
study of multi-prover quantum interactive proof systems.

The first application demonstrates that provers in a multi-prover
quantum interactive proof system may not always have an optimal
strategy when limited to finite entanglement.
We view that the primary importance of this fact is that it will serve
to better focus efforts on proving bounds on the amount of
entanglement needed for {\it close-to-optimal} provers in multi-prover
quantum interactive proofs---for such bounds can only exist in general
for close-to-optimal and not optimal success probability.

The second application is a simple proof that multi-prover quantum
interactive proof systems can be efficiently transformed to have
near-perfect completeness by adding one round of communication.
There is a trade-off between this proof and the proof of Kempe,
Kobayashi, Matsumoto, and Vidick \cite{KempeKMV09}, which is that it
is considerably simpler but cannot be said to achieve absolutely
perfect completeness.

A few other applications of coherent state exchange have also been
mentioned.  First, we have proved that the collection of
entanglement-assisted local quantum operations on systems of dimension
3 and higher is not a closed set.  Second, we have proved that
universal embezzling families exist for any number of parties.

We describe a couple of open problems.  
First, our universal embezzling families for three or more parties
appear to be highly inefficient.
Do there exist constructions that offer a significant improvement in
efficiency?
Second, as alluded to above, it is interesting to consider
near-optimal strategies for multiple provers, and it is unclear how
much entanglement is needed in such cases.
More specifically, consider all possible two-player cooperative quantum
games of the type defined in Section \ref{sec:game-description} with
fixed dimensions for $\reg{R}$, $\reg{S}$, $\reg{T}$, $\reg{A}$, and
$\reg{B}$, and fix a constant $\epsilon>0$.  
It would be interesting to find a uniform upper bound of the size of
the auxiliary entangled state such that, for each game, an approximate
strategy exists whose winning probability is $\epsilon$-close to the
optimal value.


\bibliographystyle{alpha}

\newcommand{\etalchar}[1]{$^{#1}$}

\end{document}